\documentclass[12pt,a4paper]{article}

\usepackage[british]{babel}

\usepackage[a4paper,top=2cm,bottom=2cm,left=2.5cm,right=2.5cm,marginparwidth=1.75cm]{geometry}

\usepackage[left]{lineno}

\usepackage[
    natbib=true,
    style=nature,
    sorting=none,
    eprint=false,
    maxcitenames=5,
    autocite=superscript,
    date=year,
    useeditor=false
]{biblatex}
\DeclareFieldFormat{title}{#1} 
\DeclareFieldFormat[dataset]{title}{#1.}
\DeclareSourcemap{ 
  \maps[datatype=bibtex, overwrite]{
    \map{
      \step[fieldset=editor, null]
    }
  }
}

\DeclareBibliographyDriver{online}{
  \usebibmacro{author/editor+others/translator+others}
  \setunit{\labelnamepunct}\newblock
  \usebibmacro{title}
  \newunit
  \printlist{language}
  \newunit\newblock
  \usebibmacro{byauthor}
  \newunit\newblock
  \usebibmacro{byeditor+others}
  \newunit\newblock
  \printfield{version}
  \newunit
  \printfield{note}
  \newunit\newblock
  \printlist{organization}
  \newunit\newblock
  \iftoggle{bbx:eprint}
    {\usebibmacro{eprint}}
    {}
  \newunit\newblock
  \setunit{}
  \usebibmacro{issue+date}
  \newunit\newblock
  \usebibmacro{addendum+pubstate}
  \setunit{\bibpagerefpunct}\newblock
  \usebibmacro{pageref}
  \newunit\newblock
  \iftoggle{bbx:related}
    {\usebibmacro{related:init}
     \usebibmacro{related}}
    {}
  \usebibmacro{finentry}
}

\DeclareBibliographyDriver{dataset}{
  \usebibmacro{author/editor+others/translator+others}
  \setunit{\labelnamepunct}\newblock
  \usebibmacro{title}
  \newunit\newblock
  \emph{\printlist{publisher}}
  \setunit{}
  \usebibmacro{url}
  \setunit{}
  \usebibmacro{issue+date}
  \newunit\newblock
  \usebibmacro{addendum+pubstate}
  \setunit{\bibpagerefpunct}\newblock
  \usebibmacro{pageref}
  \newunit\newblock
  \iftoggle{bbx:related}
    {\usebibmacro{related:init}
     \usebibmacro{related}}
    {}
  \usebibmacro{finentry}
}

\usepackage{amsmath}
\usepackage{bm}
\usepackage{graphicx}
\usepackage[colorlinks=true, allcolors=blue]{hyperref}
\usepackage{hyperref}
\usepackage[title]{appendix}
\usepackage{mathrsfs}
\usepackage{amsfonts}
\usepackage{booktabs} 
\usepackage{caption} 
\usepackage{threeparttable} 
\usepackage{algorithm}
\usepackage{algorithmicx}
\usepackage{algpseudocode}
\usepackage{listings}
\usepackage{enumitem}
\usepackage{chngcntr}
\usepackage{booktabs}
\usepackage{lipsum}
\usepackage{subcaption}
\usepackage{authblk}
\usepackage[T1]{fontenc}    
\usepackage{csquotes}      
\usepackage{diagbox}
\usepackage{marvosym} 

\usepackage{setspace}
\usepackage{titlesec}
\titleformat{\section} {\normalfont\Large\bfseries}{\thesection.}{1em}{}

\usepackage{float}  
\usepackage{caption} 

\newcommand{\params}{\bm{\mathrm{\theta}}}
\newcommand{\pmmmatrix}{\bm{\mathrm{M}}}
\newcommand{\data}{\bm{\mathrm{d}}}



\begin{document}

\title{Inferring three-nucleon couplings from multi-messenger neutron-star observations}

\author[1,2,\Letter,*]{Rahul~Somasundaram}
\author[3,4,5,\Letter,*]{Isak~Svensson}
\author[2]{Soumi~De}
\author[6]{Andrew~E.~Deneris}
\author[3,4]{Yannick~Dietz}
\author[7,8]{Philippe~Landry}
\author[3,4,5]{Achim~Schwenk}
\author[2]{Ingo~Tews}

\affil[1]{\small{Department of Physics, Syracuse University, Syracuse, NY 13244, USA}}
\affil[2]{Theoretical Division, Los Alamos National Laboratory, Los Alamos, NM 87545, USA}
\affil[3]{Technische Universit\"at Darmstadt, Department of Physics, 64289 Darmstadt, Germany}
\affil[4]{ExtreMe Matter Institute EMMI, GSI Helmholtzzentrum f\"ur Schwerionenforschung GmbH, 64291 Darmstadt, Germany}
\affil[5]{Max-Planck-Institut f\"ur Kernphysik, Saupfercheckweg 1, 69117 Heidelberg, Germany}
\affil[6]{\small{Computer, Computational, and Statistical Sciences Division, Los Alamos National Laboratory, Los Alamos, NM 87545, USA}}
\affil[7]{Canadian Institute for Theoretical Astrophysics, University of Toronto, Toronto, Ontario M5S 3H8, Canada}
\affil[8]{Perimeter Institute for Theoretical Physics, Waterloo, Ontario N2L 2Y5, Canada}
\affil[*]{Equally contributing authors}
\affil[\Letter]{E-mail: rsomasundaram@lanl.gov; isak.svensson@tu-darmstadt.de}

\date{}  

\maketitle

\begin{abstract}
Understanding the interactions between nucleons in dense matter is an important challenge in theoretical physics. 
Effective field theories have emerged as the dominant approach to address this problem at low energies, with many successful applications to the structure of nuclei and the properties of dense nucleonic matter. 
However, how far into the interior of neutron stars these interactions can describe dense matter is an open question.
Here, we develop a framework that enables the inference of three-nucleon couplings in dense matter directly from astrophysical neutron star observations. 
We apply this formalism to the LIGO/Virgo gravitational-wave event GW170817 and the X-ray measurements from NASA's Neutron Star Interior Composition Explorer and establish direct constraints for the couplings that govern three-nucleon interactions in chiral effective field theory. Furthermore, we demonstrate how next-generation observations of a population of neutron star mergers can offer stringent constraints on three-nucleon couplings, potentially at a level comparable to those from laboratory data. 
Our work directly connects the microscopic couplings in quantum field theories to macroscopic observations of neutron stars, providing a way to test the consistency between low-energy couplings inferred from terrestrial and astrophysical data.
\end{abstract}

Obtaining a description of strong-interaction matter based on the underlying theory of quantum chromodynamics (QCD) is one of the most important challenges of theoretical physics. 
The development of effective field theories (EFTs) of QCD has revolutionized our understanding of nuclear forces~\autocite{Weinberg:1990rz,Epelbaum:2008ga,Hammer:2019poc}.
By employing a separation of scales between typical nucleon momenta and high-energy degrees of freedom, indicated by the breakdown scale of the theory, EFTs provide a systematic expansion for inter-nucleon interactions which can be improved order by order. 
Consequently, a natural hierarchy emerges among the interactions between two (NN), three (3N), and many nucleons, the strengths of which are parameterized by low-energy couplings (LECs) that need to be determined from data or from the underlying theory. 
In the absence of direct calculations of the LECs from QCD, the LECs are commonly adjusted to reproduce nuclear experiments.
For example, LECs that govern NN interactions are fit to NN scattering data whereas 3N forces are fit to properties of light nuclei.  The LECs governing the two-pion exchange contribution to the 3N forces~\autocite{Epelbaum:2002vt}, $c_1$ and $c_3$, describe pion-nucleon ($\pi$N) interactions and can be constrained using laboratory $\pi$N scattering data to $c_1=-0.74\pm 0.02$~GeV$^{-1}$ and $c_3=-3.61 \pm 0.05$~GeV$^{-1}$ at one-sigma level; see Hoferichter et al.~\autocite{Hoferichter:2015hva} and Siemens et al.\autocite{Siemens:2016jwj} for an analysis using Roy-Steiner equations.
EFT-based interactions with LECs calibrated this way have enjoyed significant success in applications to nuclear structure and nuclear astrophysics~\autocite{Hergert:2020bxy,Hebeler:2020ocj,Hagen:2015yea,Gysbers:2019uyb,Stroberg:2019bch,Hu:2021trw,Hebeler:2013nza,Lynn:2015jua,Tews:2018iwm,Drischler:2017wtt,Drischler:2021kxf,Keller:2022crb}.
However, an open question is which data with $A \geq 3$ one should use to infer LECs describing three-nucleon interactions.
For example, recently the calibration of certain LECs to medium-mass nuclei, such as oxygen isotopes, has resulted in increased success for nuclear structure phenomenology, enabling EFT-based Hamiltonians to successfully describe nuclei up to $^{208}$Pb (see, e.g., Hu et al.\autocite{Hu:2021trw} and Arthuis et al.\autocite{Arthuis:2024mnl}). 
These developments emphasize that it is key to identify critical experimental constraints for the calibration of powerful nuclear interactions.

Interestingly, the microscopic interactions that govern atomic nuclei also dictate the macroscopic properties of neutron stars.
Neutron stars are fascinating systems that access higher densities and greater neutron excess than nuclei accessible in terrestrial experiments. 
Therefore, they offer an unparalleled laboratory for nuclear matter in regions of high density and isospin asymmetry (the ratio of neutrons to protons)~\autocite{Capano:2019eae,Dietrich:2020efo,Al-Mamun:2020vzu,Essick:2021kjb,Huth:2021bsp,Annala:2023cwx,Raaijmakers:2021uju,Rutherford:2024srk}. 
However, significant challenges arise in modeling the equation of state (EOS) of neutron stars, such as the involved computational cost of solving the quantum many-body problem. 
Initial steps have been taken to constrain 3N interactions from astrophysical observations~\autocite{Maselli:2020uol,Sabatucci:2022qyi,Rose:2023uui} but none of these studies have attempted to calibrate the nuclear Hamiltonian directly to such data. 
While Maselli et al.~\autocite{Maselli:2020uol} and Sabatucci et al.~\autocite{Sabatucci:2022qyi} explore the sensitivity of neutron star observables to a single parameter governing the strength of the short-range 3N contribution to the EOS, they do not solve the quantum many-body problem at each step in the sampling but instead make simplifying assumptions to connect the single parameter to neutron star observables.
Furthermore, these studies do not fully model variations in the EOS from low to high densities, beyond the nuclear regime, where the density dependence can change from that dominated by 3N forces. For example, phase transitions to non-nuclear degrees of freedom are possible. 
Rose et al.~\autocite{Rose:2023uui} use astrophysical observations to distinguish between two Hamiltonians that model the 3N force in different ways but do not infer the LECs themselves.

Here, we develop a framework that allows us to constrain LECs directly from observations of neutron stars via Bayesian inference and, thus, explore EFT-based interactions for the densest neutron-rich systems in the cosmos. 
We consider a Hamiltonian at next-to-next-to-leading order (N$^2$LO) in the EFT expansion and focus on the leading (N$^2$LO) 3N forces that provide a strong contribution to the EOS of neutron matter. 
As LEC values obtained from two entirely distinct sources (neutron star observations versus atomic nuclei) can be checked against each other for consistency, our framework will provide a unique test for the domain of applicability of nuclear interactions and of the convergence of the EFT expansion in dense matter. \\

\subsection*{Results}

\noindent \textbf{Machine-learning--based inference framework} 

\begin{figure}
    \centering
    \includegraphics[width=0.99\linewidth]{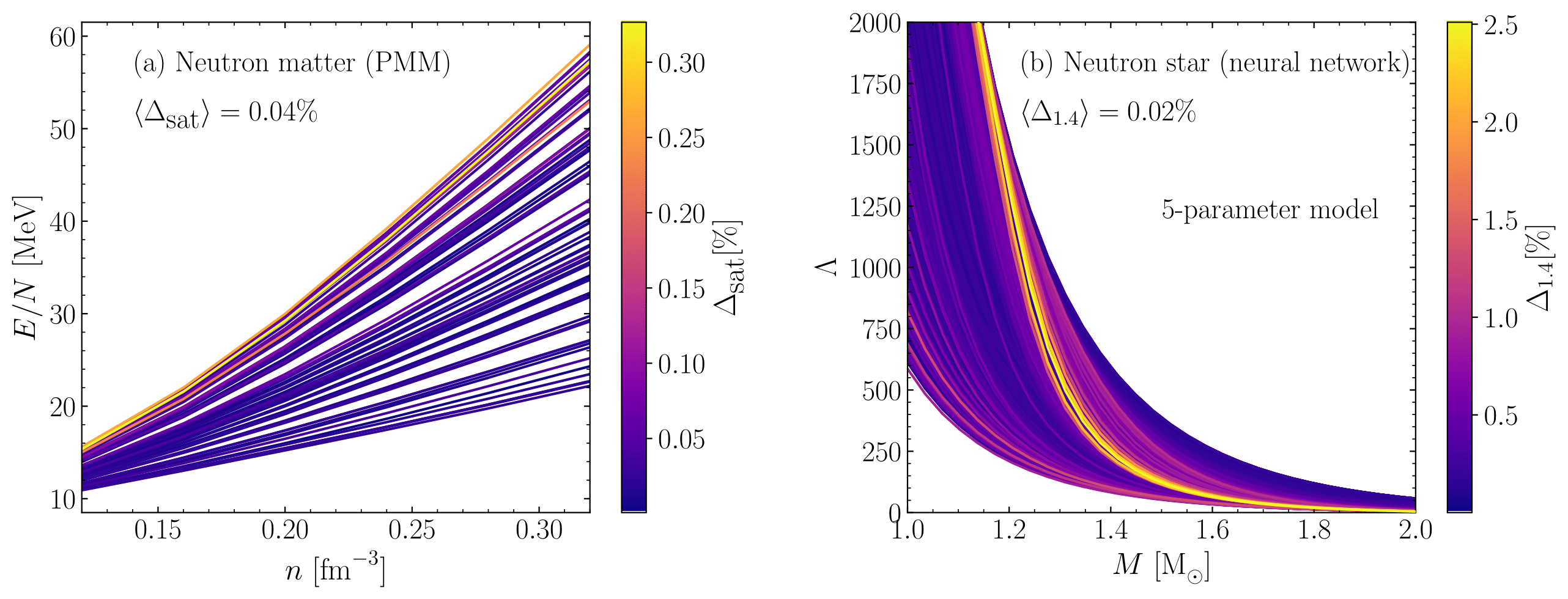}
    \caption{\textbf{Validation results for our machine-learning--based emulators}. 
    In both panels, the color of the curves corresponds to the percentage uncertainty in the emulators' prediction. 
    Panel (a) shows $70$ results for the neutron-matter EOS up to $2n_\textrm{sat}$ predicted by a PMM built using a training set of $30$ different samples.
    At nuclear saturation density, individual validation uncertainties $\Delta_{\rm sat}$ are below 0.3\%, with the average uncertainty being 0.04\%.
    In Panel (b), we show about 60,000 tidal deformability-mass curves predicted by an ensemble of 50 neural networks trained on the 5-parameter EOS model.
    The uncertainty at $1.4 M_{\odot}$, $\Delta_{1.4}$, is at most 2.5 \% with an average of 0.02 \%. 
    We note that less than 0.1\% of EOSs have emulator uncertainties of around 2\%, making these very rare occurrences. Source data for this figure are provided as a Source Data file.
    }
    \label{fig:emulators}
\end{figure}

Our Bayesian inference setup to constrain $c_1$ and $c_3$ from astrophysical data involves sampling over the microphysical LECs using a Markov chain Monte Carlo (MCMC) stochastic sampling algorithm.
Every iteration of the sampler requires the solution of the many-body Schr{\"o}dinger equation, which yields the neutron star EOS.
The EOS is subsequently translated to astrophysical observables, such as radii or tidal deformabilities, by solving the Tolman-Oppenheimer-Volkoff (TOV) equations and the equations for a stationary quadrupolar tidal deformation~\autocite{Flanagan:2007ix,Hinderer:2009ca}.
Given the large number of iterations required to sample the posterior distribution function---$\mathcal{O}(10^6)$---this is a computationally intractable problem given that a single iteration requires $\mathcal{O}(10^2)$ CPU-h. 

We overcome this challenge by employing recent advances in machine-learning--based algorithms that act as surrogate models to more complex high-fidelity calculations.  
First, for rapid calculation of the EOS, we employ the recently proposed parametric matrix model (PMM)~\autocite{Cook:2024toj,Somasundaram:2024zse} which is trained on third-order many-body perturbation theory (MBPT) calculations~\autocite{Keller:2022crb} of the neutron-matter EOS. 
In the left panel of Fig.~\ref{fig:emulators}, we show results for $70$ validation samples for a PMM trained on $30$ high-fidelity MBPT calculations. 
We find emulator uncertainties to be well under control, with predictions of the EOS at nuclear saturation density $n_{\rm sat}=0.16$~fm$^{-3}$ differing from the high-fidelity results by 0.04\% on average.
This emulator uncertainty is much smaller than the LEC variation, indicated by the spread of samples, as well as the uncertainty in the MBPT calculations themselves (see Methods).
We then extend these neutron-matter results to neutron star matter in beta equilibrium.
We model the neutron star EOS using three different parameterizations (see Methods), the first and simplest of which only uses the emulator results based on $c_1$ and $c_3$ to characterize the EOS up to $10 n_\textrm{sat}$.

The nucleonic description of dense matter eventually breaks down and exotic phenomena such as QCD phase transitions in high-density matter might appear. 
Therefore, in the following we assume the validity of the EFT expansion up to $2 n_\textrm{sat}$~\autocite{Tews:2018kmu,Drischler:2020hwi,Keller:2022crb}, and hence, the LECs $c_1$ and $c_3$ determine the EOS only up to this density. 
An important aspect of our framework is to find a suitable parameterization and marginalization over uncertainties in the high-density EOS. 
Here, we use two different models based on the speed of sound that allow for a physics-agnostic extension of the EOS above $2 n_\textrm{sat}$~\autocite{Somasundaram:2021clp}.
These models employ either 5 or 7 parameters including the two LECs. 
Comparing these models with the 2-parameter model illustrates the importance of the marginalization over the high-density EOS.

For a given EOS model, we use an ensemble of neural networks to predict the tidal deformability of neutron stars~\autocite{Reed:2024urq}. 
Results obtained from validating our emulator on about 60,000 samples are shown in the right panel of Fig.~\ref{fig:emulators} for the 5-parameter model, but the emulator performance is similar for the other EOS models.
We find that both the PMM and neural network emulators provide highly accurate and rapid emulation of the underlying complex calculations that is required to compute neutron star properties starting from microscopic LECs. 
The use of machine-learning algorithms in this manner is a key part of our analysis that allows us to sample complex posterior distribution functions in our astrophysical Bayesian analysis framework. \\

\noindent \textbf{Inference based on GW170817 and NICER data}

\begin{figure}[t]
    \centering
    \includegraphics[width=0.65\textwidth]{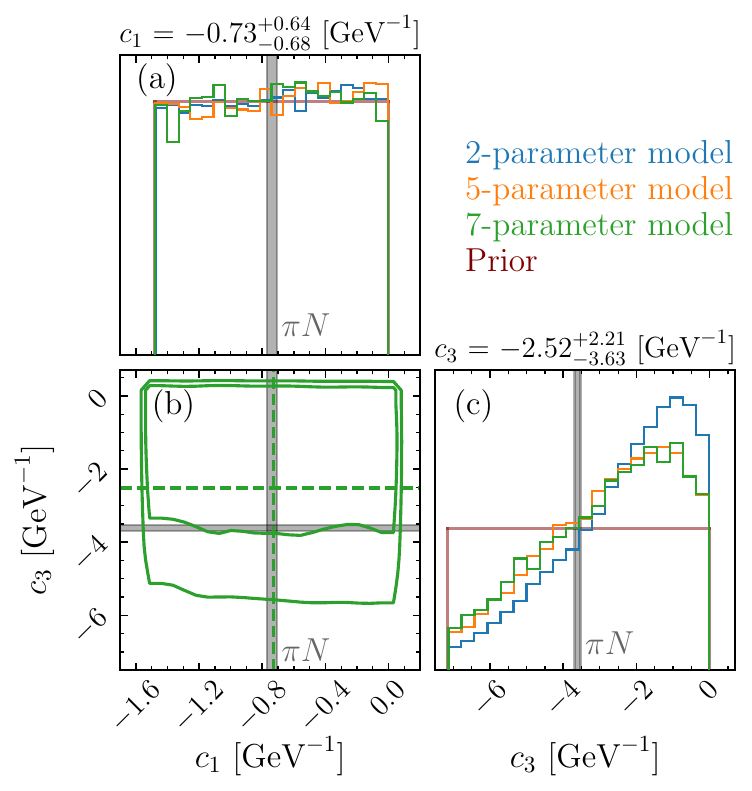}
    \caption{\textbf{Constraints from GW170817 and NICER data}. 
    Panel (a) shows the posterior distribution function of the LEC $c_1$. Panel (c) shows the posterior distribution function of the LEC $c_3$. 
    In panels (a) and (c), results are shown for the 2- (blue), 5- (orange), and 7-parameter (green) models. 
    The quoted errors are for the 7-parameter model evaluated at the 90$\%$ confidence level. 
    Panel (b) depicts the correlation between $c_1$ and $c_3$ and displays iso-probability contours at the $68\%$ and $90\%$ confidence levels. 
    The laboratory values for $c_1$ and $c_3$ from pion-nucleon ($\pi N$) scattering experiments are shown in gray. Source data for this figure are provided as a Source Data file.}
    \label{fig:Present}
\end{figure}

We condition the LECs $c_1$ and $c_3$ on the LIGO/Virgo Collaboration's first GW observation of a binary neutron star merger, GW170817~\autocite{LIGOScientific:2017vwq,LIGOScientific:2018hze}, as well as X-ray observations of three pulsars made by NASA's Neutron Star Interior Composition Explorer (NICER) mission: PSR J0030+0451~\autocite{Riley:2019yda,Miller:2019cac}, PSR J0740+6620~\autocite{Riley:2021pdl,Miller:2021qha}, and PSR J0437--4715~\autocite{Choudhury:2024xbk}. 
For these observations, we employ likelihood functions of the source parameters (neutron star masses, binary tidal deformabilities, radii, etc.) that are proportional to the posteriors computed in Abbott et al.\autocite{LIGOScientific:2018hze} for GW170817, and Riley et al.\autocite{Riley:2019yda}, Salmi et al.\autocite{Salmi:2022cgy}, and Choudhury et al.\autocite{Choudhury:2024xbk} for the three NICER observations.   
With wide uniform priors between zero and twice the mean of their laboratory values on the LECs $c_1$ and $c_3$, the posterior conditioned on GW170817 is sampled using MCMC, and the posterior samples are weighted according to the product of the NICER likelihoods.
Each likelihood evaluation is carried out by evaluating our two emulators consecutively, thus, converting microscopic LECs to macroscopic neutron star observables.

The marginalized two-dimensional posterior distribution on $c_1$ and $c_3$ is shown in Fig.~\ref{fig:Present}. 
We find essentially no influence of the neutron star observations on the parameter $c_1$ since the effect of $c_1$ on the EOS of neutron matter is subdominant compared to that of $c_3$~\autocite{Hebeler:2009iv}.
On the other hand, given the GW and X-ray observations, we see a clear preference for less repulsive 3N forces.
We attribute this to $c_3$ being negatively correlated with the pressure inside neutron stars~\autocite{Hebeler:2013nza}.
Since GW170817 as well as the X-ray observation of PSR J0470 favor more compact neutron stars, this translates to smaller pressures in neutron star interiors.
The upper bound of the posterior is set by the prior boundary at $c_3=0$~GeV$^{-1}$; however, this boundary is physically motivated since positive values of $c_3$ correspond to an attractive 3N force, which would lead to collapse of neutron matter. 
Although the posterior median of $c_3=-2.52$~GeV$^{-1}$ deviates from the laboratory value of $c_3 = -3.61\pm 0.05$~GeV$^{-1}$, the laboratory and astrophysical determinations are consistent at the 90\% confidence level, albeit with the latter constraint currently having large uncertainties.
Existing neutron star observations do not offer high-precision constraints on the LECs given the significant statistical uncertainties present in these observations.\\

\noindent \textbf{Constraints from next-generation GW observatories}
 
In the next decade, two upcoming next-generation ground-based GW detectors are expected to begin operations, namely the Einstein Telescope (ET) in Europe~\autocite{Punturo:2010zz,Maggiore:2019uih} and Cosmic Explorer (CE) in the United States~\autocite{Reitze:2019iox,Evans:2021gyd}.
These detectors will provide a sensitivity that is improved by an order of magnitude over current GW detectors.
Consequently, they are expected to observe $\mathcal{O}(100)$ events with signal-to-noise (SNR) ratios above $100$ per year~\autocite{Borhanian:2022czq,EvansCorsi2023,BranchesiMaggiore2023,Gupta:2023lga,Smith:2021bqc,Finstad:2022oni}. 
Here, we demonstrate how such observations, at the level of populations of events, can provide stringent constraints on 3N couplings.

\begin{figure}
    \centering
    \includegraphics[width=0.89\linewidth]{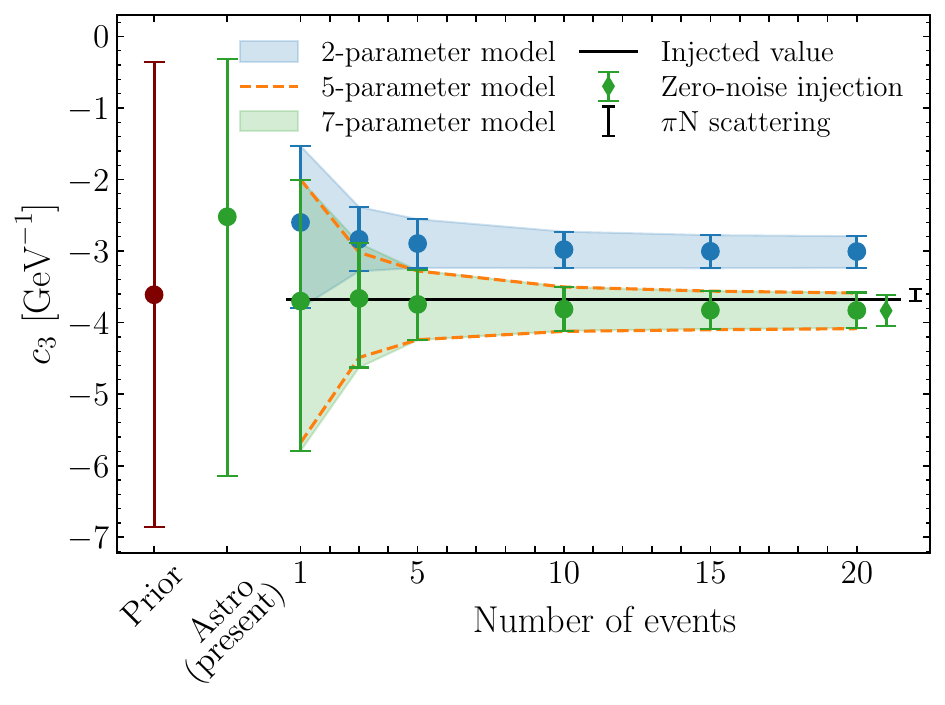}
    \caption{\textbf{Evolution of the obtained constraints on $c_3$}. 
    We show the prior on $c_3$ (purple) as well as the present astrophysical constraints, which include GW170817 and the NICER X-ray observations of millisecond pulsars but exclude simulated data from next-generation GW observatories. 
    The inferred $c_3$ for our population study is then shown as a function of the number of events observed by a network of next-generation GW detectors for the 2-parameter (blue), 5-parameter (orange-dashed) and 7-parameter (green) EOS models.
    We observe that the constraints are well converged at 20 observations.
    For the 7-parameter model, the diamond error bar is obtained by analyzing 20 observations using parameter estimation on zero-noise injections, whereas the circle error-bars are obtained within a Fisher matrix approximation. 
    We compare the resulting constraints to the injected value and to the uncertainties in the laboratory result from $\pi$N scattering (black). All error bars represent $90\%$ confidence levels. Source data for this figure are provided as a Source Data file.}
    \label{fig:convergence}
\end{figure}

We simulate a population of neutron star merger events that can potentially be observed within a year-long observing run by a network of three next-generation detectors. 
Our detector network consists of two CEs, each with a 40~km arm length, and one ET with its fiducial 10~km triangular design. 
For our population model, we assume a uniform distribution in the range 1--2~$M_\odot$ for the component masses and a random pairing into binary systems. 
Furthermore, we assume a constant local merger rate of $170$ Gpc$^{-3}$ yr$^{-1}$, which is consistent with the merger rate inferred in Abbott et al.\autocite{KAGRA:2021duu}, and a uniform distribution of sources in co-moving volume.
This results in a total of approximately 400 events within a redshift $z$ of $0.2$ observed by the detector network within one year. 
A lower merger rate would only affect our results by increasing the time required to observe the same number of events.
There are many additional events with $z > 0.2$ that are detected with lower SNR; however, here we focus on the loudest signals in the detected population.
The underlying EOS explored by our population, i.e., the EOS that relates the component masses of our simulated binaries to the corresponding component tidal deformabilities, is generated from our 5-parameter model with $c_3 = -3.68$~GeV$^{-1}$, which is in agreement with the laboratory value at the 90\% confidence level~\autocite{Siemens:2016jwj}.    

From the detected events in our simulated population, we select the $N=20$ highest-SNR events to analyze. This most informative subset is large enough that its chirp masses span the entire range allowed by the population model, with at least one event in every chirp mass bin of width $0.1 M_{\odot}$. This choice amounts to a selection cut at an SNR of about 225. Because the SNR depends primarily on the chirp mass, whose distribution is independent of the EOS in our assumed population model, this selection cut does not bias the recovery of EOS parameters. GW selection effects can thus be neglected in our inference.

The evolution of the uncertainties in $c_3$ as a function of the number $N$ of observed events, represented at the $90\%$ confidence level, is shown in Fig.~\ref{fig:convergence}.
The $N$ events are analyzed using a Bayesian hierarchical approach~\autocite{LandryEssick2020,Landry:2022rxu} for $N=1,3 ,5, 10, 15$ and $20$, within the Fisher matrix approximation~\autocite{Cutler:1994ys,Poisson:1995ef}.
For $N=20$, the analysis is repeated using parameter estimation on zero-noise injections~\autocite{Veitch:2014wba,Biwer:2018osg}. The zero-noise realization of Gaussian noise is, in the statistical sense, the most likely realization~\autocite{Smith:2021bqc} and, furthermore, the effect of any non-zero Gaussian noise generally weakens drastically with the SNR.   
The result matches well with the corresponding Fisher matrix result (see Figure~\ref{fig:convergence}). This is consistent with other findings in the literature, see for example Vallisneri~\autocite{Vallisneri:2007ev}, which demonstrate that, for a four dimensional Fisher matrix analysis, an SNR above 20 is typically sufficient for robust parameter estimation. 
We find that a single event among the 20 loudest ones observed by next-generation GW detectors in a year decreases uncertainties by only, on average, a factor two compared with uncertainties obtained from present astrophysical data. 
However, as the number of detections increases, we find that the statistical uncertainties in $c_3$ decrease approximately as $1/\sqrt{N}$, and thus converge remarkably well to the injected value. 
We see how inference performed at the level of populations of events can potentially provide high-precision constraints on nuclear interactions within a year, competitive with and complementary to terrestrial laboratory data.
These constraints would improve further if events with lower SNR are also considered. In contrast, the constraints could weaken in the same time frame if, for example, the observed merger rate is lower than expected.

Figure~\ref{fig:convergence} also demonstrates the importance of the marginalization over uncertainties in the high-density EOS, implemented in our framework in the 5- and 7-parameter models.
We find excellent agreement between results obtained from these two models. 
However, the very simple 2-parameter model---which does not account for such uncertainties---is in significant tension with the injected value, as it converges toward an incorrect $c_3$. 
This underscores the importance of allowing for general high-density extensions in order to avoid systematic uncertainties in the inference of LECs. \\

\noindent \textbf{Discussion}

In this paper we introduce a framework that allows for the inference of couplings describing microscopic 3N interactions from astrophysical observations of neutron stars.
Naively, one would assume that the complex multi-physics calculations required to compute neutron star observables starting from microscopic Hamiltonians are too computationally expensive to allow for a full stochastic sampling of the posterior distribution. 
We have overcome this challenge by enhancing our Bayesian inference approach with machine learning.
In particular, we employ two machine learning methods---the parametric matrix model~\autocite{Cook:2024toj,Somasundaram:2024zse} and the ensemble neural-network method\autocite{Reed:2024urq,scikit-learn}---that circumvent this challenge by drastically speeding up our likelihood evaluations.

The inference of LECs from neutron star observations provides several benefits.
While LECs can be adjusted to data on atomic nuclei or scattering, our novel approach makes it possible to constrain interactions using the densest and most neutron-rich system in the cosmos.
Furthermore, it is possible that improved nuclear interactions involve not only higher orders in the EFT expansion, but the addition of new degrees of freedom, such as the $\Delta$ resonance~\autocite{Jiang:2020the} or hyperons~\autocite{Petschauer:2020urh}. 
Even existing EFTs that include such degrees of freedom suffer from a lack of experimental data for its calibration and verification. 
For all these cases, astrophysical data provides a viable option for the calibration of the LECs. 

While we have chosen to study only the 3N LECs $c_1$ and $c_3$, our framework is applicable to any sector of the Hamiltonian provided that it correlates sufficiently well with the EOS of neutron-rich matter. 
This enables consistency checks between constraints from neutron star observations and from nuclear experiments, which provides a nontrivial verification of the EFT expansion and its applicability to calculations of neutron star observables.
Our framework also enables sensitivity analyses of neutron-rich matter that will allow for the identification of the most important LECs in chiral Hamiltonians for neutron star observables. 
In Fig.~\ref{fig:convergence}, the extracted value of $c_3$ is in good agreement with laboratory data by choice. 
Such a result with actual GW data would demonstrate consistency of the EFT approach between nuclei and dense matter.
However, a hypothetical convergence of $c_3$ to a value incompatible with laboratory data---similar to the result for the 2-parameter model---might indicate the breakdown of the EFT and the appearance of new physics in neutron stars.
Our framework also permits us to compare different implementations of EFTs as well as various many-body methods for strongly correlated quantum systems. 
With the advent of next-generation neutron star observations in the near future, we therefore expect our framework to be a key tool for inferring LECs and improvements to the EFTs themselves.
Our work thus also demonstrates the importance of building a global network of next-generation ground-based gravitational-wave observatories.

\subsection*{Methods}

\noindent \textbf{High-fidelity EOS model}

We require EOS models, connecting the nuclear Hamiltonian to neutron star structure, to generate training data for the machine learning-based algorithms employed in our Bayesian inference framework. 
In this work, we keep the NN part of the Hamiltonian fixed using the N$^2$LO NN interactions of Entem, Machleidt, and Nosyk (EMN) with cutoff $\Lambda_c = 450$~MeV~\autocite{Entem:2017gor}, as this part is very well determined from NN scattering experiments.
At N$^2$LO, a given nuclear Hamiltonian for neutron matter is then parameterized by the values of the LECs $c_1$ and $c_3$.
To generate the EOS from this Hamiltonian, the first step is to solve the many-body Schr{\"o}dinger equation.
Here, we use MBPT at third order~\autocite{Drischler:2017wtt,Keller:2022crb} to calculate the energy per particle of neutron matter between $0.12$~fm$^{-3}$ and $0.32$~fm$^{-3}$ in steps of $0.04$~fm$^{-3}$ for 100 randomly selected sets of LEC combinations.
For symmetric nuclear matter, which is not as important but is needed to include the small fraction of protons, we use empirical properties.
We estimate the uncertainty of the MBPT approach by comparing the difference of results at second and third order.
The average (maximum) size of the third-order MBPT correction is 252 (790) keV across all densities and for all LEC combinations, which is much smaller than the overall range of the energy per particle; see Fig.~\ref{fig:emulators}.
The correction generally increases with density away from saturation density.
Based on this we conclude that the MBPT calculations are sufficiently converged so as to not significantly impact our results.
For this work, we use an MBPT code that is based on 3N operators in a single-particle basis, as in Drischler et al.~\autocite{Drischler:2017wtt}. 
This makes it computationally feasible to perform the calculations required for training our emulators.

Using an MBPT calculation of the EOS of neutron matter, we compute the EOS for neutron star matter in beta equilibrium up to $2n_\textrm{sat}$ ($10n_\textrm{sat}$) using the metamodel of Margueron et al.\autocite{Margueron:2017eqc,Margueron:2017lup} for our 5- and 7- (2-) parameter models. 
The metamodel provides a smooth density-functional--based interpolation between the discrete density-grid points of the MBPT calculation and allows for a straightforward extrapolation from the EOS of neutron matter to the EOS of matter in beta equilibrium including leptons. 
The model parameters of the metamodel are the so-called nuclear empirical parameters (NEPs) that govern the density dependence of the EOSs of neutron matter and symmetric matter via a Taylor expansion about nuclear saturation density, see Margueron et al.~\autocite{Margueron:2017eqc} for more details. 
Here, we fix the iso-scalar NEPs to be $E_\textrm{sat} = -16$~MeV, $n_\textrm{sat} = 0.16$~fm$^{-3}$, $K_\textrm{sat} = 230$~MeV, and $Q_\textrm{sat} = Z_\textrm{sat}  = 0$~MeV, i.e., we do not consider variations in the EOS of symmetric matter in this work. 
On the other hand, the iso-vector parameters are fit to the MBPT calculation of the energy per particle up to $2n_\textrm{sat}$ in neutron matter. 
For the 5- and 7-parameter models, the fit includes the iso-vector parameters $E_\textrm{sym}$, $L_\textrm{sym}$, $K_\textrm{sym}$, and $Q_\textrm{sym}$, i.e., all the iso-vector NEPs excluding $Z_\textrm{sym}$ which is fixed at $0$~MeV, in order to reliably reproduce the MBPT calculation. 
We found that the difference between the MBPT calculation and the metamodel result, averaged over density and different LEC samples, is about 10~keV.
For the 2-parameter model, the NEP $Q_\textrm{sym}$ is fixed to $Q_\textrm{sym}=500$~MeV, which results in a slight loss in the quality of the fit but allows for an extension of the neutron-star EOS up to $10n_\textrm{sat}$ using the metamodel alone while ensuring that the symmetry energy remains positive up to this density. 
Finally, for all three of our EOS models, the crust EOS of Douchin and Haensel~\autocite{Douchin:2001sv} is added below 0.08~fm$^{-3}$ following the approach of Koehn et al.~\autocite{Koehn:2024set}.

For the 5- and 7-parameter models, we account for phenomena beyond the EFT breakdown scale---such as the appearance of non-nucleonic degrees of freedom at high densities---by extending the EOS above $2n_\textrm{sat}$ using the speed-of-sound parameterization~\autocite{Tews:2018iwm,Somasundaram:2021clp}.
This parameterization is agnostic towards the composition of matter; the model parameters are the values of the squared speed of sound, defined on a discrete density grid. 
For the 5- (7-) parameter model, we use three (five) grid points located at $[3,5,7]\, n_\textrm{sat}$ ($[3,4,5,6,7]\, n_\textrm{sat}$).
Then, for a given set of sound speeds on the specified grid, the full sound-speed curve is obtained by interpolating linearly between the grid points.  
In both models, the speed of sound at $n > 7 n_\textrm{sat}$ is taken to be a constant equal to its value at $7 n_\textrm{sat}$. The full density-dependent speed of sound can then be integrated to obtain the pressure, energy density and baryon chemical potential~\autocite{Tews:2018iwm,Somasundaram:2021clp}.

Using the neutron star EOS generated in this manner, we obtain the neutron star radius $R$ and tidal deformability $\Lambda$ as a function of the neutron star mass by solving the TOV equations and the equation for the quadrupolar tidal perturbation~\autocite{Flanagan:2007ix,Hinderer:2009ca}.\\

\noindent \textbf{Machine learning the EOS model}

The sequence of steps which is required to convert a set of EOS parameters to neutron-star observables, such as the tidal deformability, is computationally expensive and requires $\mathcal{O}(10^2)$~CPU-h for a single evaluation. 
This is prohibitively expensive given that the total number of iterations required in this work is $\mathcal{O}(10^9)$. 
We overcome this bottleneck by using machine-learning approaches to emulate the two steps necessary to obtain the high-fidelity EOS model and connect it to neutron stars.

The first and most expensive step of generating an EOS model is the computation of the neutron matter EOS using MBPT.
To emulate this computation, we use the PMM~\autocite{Cook:2024toj,Somasundaram:2024zse} that combines ideas of both machine learning and reduced-order modeling. 
In our implementation of the PMM, the energy per particle in neutron matter at a given density is represented by the lowest eigenvalue of a $2 \times 2$ matrix $\pmmmatrix$ given as
\begin{equation}
    \pmmmatrix = \pmmmatrix_0 + c_1 \pmmmatrix_1 + c_3 \pmmmatrix_3\,,
\end{equation}
where $\pmmmatrix_0$ is a diagonal matrix, $\pmmmatrix_1$ and $\pmmmatrix_3$ are symmetric matrices, and $c_1$ and $c_3$ are the two LECs of interest. 
While the dependence of $\pmmmatrix$ is inspired by the theory of reduced-order modeling and the affine structure of chiral Hamiltonians, the matrix elements of $\pmmmatrix_0$, $\pmmmatrix_1$, and $\pmmmatrix_3$ need to be learned from data.
To achieve this, we sample $100$ different LEC combinations from their uniform prior distribution and calculate the EOS of neutron matter using MBPT. 
Then, we divide the set of $100$ samples into $30$ training and $70$ validation samples, and fit the real matrix elements of $\pmmmatrix_0$, $\pmmmatrix_1$, and $\pmmmatrix_3$ at each density such that the lowest eigenvalues of $\pmmmatrix$ agree with the MBPT results for the training set.  
Results from validating the PMM on the remaining $70$ samples are shown in Fig.~\ref{fig:emulators}. 
To verify the accuracy of our PMM in the posterior parameter space, we averaged the emulator uncertainty using only those samples that lie within the span of the posterior on $c_3$ and found an average of $0.03$\%.
The average (maximum) deviation between the PMM and the MBPT calculation, averaged over density and the 70 validation samples, is $15$~keV ($158$~keV). 

The second step is to solve the TOV equations and the equation for the quadrupolar tidal perturbation in order to calculate neutron star observables.
This process takes only $\mathcal{O}(1)$ seconds for each EOS sample.
This computational cost, while far cheaper than the MBPT calculations, 
remains a significant hurdle in Bayesian inference frameworks.
To overcome the hurdle we follow the approach of Reed et al.\autocite{Reed:2024urq} and use an ensemble of feed-forward neural networks\autocite{scikit-learn} to emulate solutions to the stellar structure equations. 
To generate training and validation data, we first sample a set of 200,000 model parameters from their uniform priors and generate the neutron star EOS for each EOS model.
During this process, we employ the PMM to calculate the corresponding EOSs of neutron matter for the various draws for $c_1$ and $c_3$ and use the meta- and sound-speed models for the neutron star EOSs.
For each of the 200,000 EOS models, we solve the TOV equations using a high-fidelity solver. 
The resulting data is then used to train the neural-network emulator, since each sample corresponds to a well-defined input (the EOS model parameters) and output (the neutron star tidal deformabilities).
In this work, the neural-network emulator is only used to predict the $\Lambda$-$M$ curve and not the $M$-$R$ curve. The latter does not require an emulator as we perform significantly fewer calculations of this quantity.

We evenly split our set of 200,000 samples into training and validation data and  discard all samples that do not reach $2$~M$_\odot$, i.e., the neural network emulator is trained to predict the $\Lambda$-$M$ curve only for those EOS parameters for which the maximum mass $M_\textrm{TOV} \geq 2 M_\odot$.
For all EOS models, the network architecture is such that it accepts the EOS parameters as input via the input layer, and then outputs the $\Lambda$-$M$ curve on a grid of 30 points uniformly spaced between 1--2 M$_\odot$.
We use five hidden layers to emulate the 5- and 7-parameter models, whereas we use two hidden layers for the 2-parameter model. All hidden layers consist of 64 nodes. 
Similar to Reed et al.\autocite{Reed:2024urq}, we use an ensemble of $50$ ($5$) neural networks for the 5- and 7-parameter (2-parameter) models, with all neural networks trained independently and the final output obtained by averaging over all neural networks in the ensemble. 

\begin{figure}[t]
    \centering
    \includegraphics[scale=0.56]{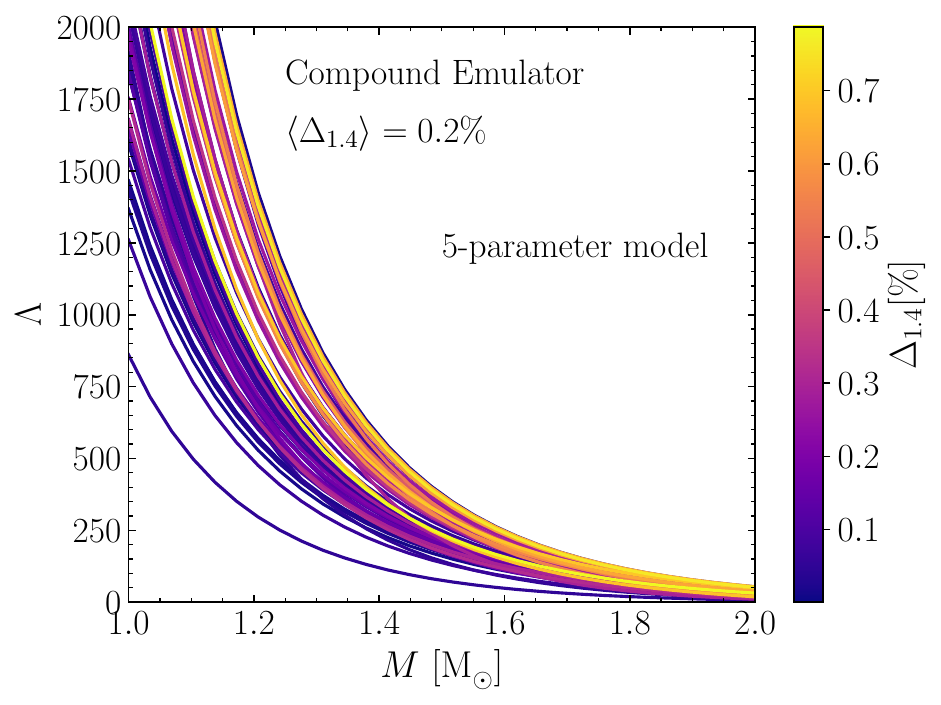}
    \caption{\textbf{Compound emulator uncertainty}. The figure shows the accuracy of our emulators where the validation is performed using both high-fidelity solvers: MBPT and the full TOV solver. 
    The quoted average error for $1.4 M_{\odot}$, $\langle\Delta_{1.4}\rangle = 0.2\%$, is an estimate of the total compound emulator uncertainty including both the PMM and the neural network, instead of only the latter. Source data for this figure are provided as a Source Data file.}
    \label{fig:compound_unc}
\end{figure}

Validation results for our neural network-based emulation are shown in the right panel of Fig.~\ref{fig:emulators}. 
However, since the validation data is computed using the PMM emulator and not high-fidelity MBPT calculations, the results of Fig.~\ref{fig:emulators} capture only the accuracy of the neural-network emulator alone and not the full compound emulation uncertainty that also includes the PMM. 
Therefore, in Fig.~\ref{fig:compound_unc}, we have computed the compound uncertainty using the 70 MBPT samples that were not used when training the PMM. 
Each of these samples was extended to higher densities using the 5-parameter model and the $\Lambda-M$ sequence was subsequently obtained using our high-fidelity TOV solver. 
We have checked that our results do not change, if the 2- or 7-parameter models are used instead. 
The error in the predicted tidal deformability at $1.4 M_{\odot}$, $\Delta_{1.4}$, while somewhat larger than in Fig.~\ref{fig:emulators}, is sufficiently small for our machine-learning-based analyses to remain unbiased. Quantitatively, an event with an SNR of about 4000---far higher than any event considered here---would be required for the average compound error of 0.2\% to be detectable with 90\% confidence.

Since the emulator is trained on only those EOS parameter values that satisfy $M_\textrm{TOV} \geq 2 M_\odot$, and therefore capable of making predictions only for those samples, we require an additional machine-learning tool, trained and validated on the same sets as above, to identify whether a sampled set of EOS parameters satisfies this criterion. 
We train an ensemble of ten neural networks to perform this binary classification, where each network accepts the EOS parameters as input and outputs the probability that the required constraint is satisfied.
The final classification is obtained by averaging over the outputs of all ten neural networks. 
We found that for the 5-parameter and 7-parameter models, our classifier successfully identifies samples that satisfy $M_\textrm{TOV} \geq 2 M_\odot$ with an accuracy of $99.9\%$. 
In the very few cases where the classifier fails, the maximum mass is usually very close to the 2 $M_\odot$ threshold and the emulator nevertheless provides a reasonable emulation.
The 2-parameter model does not require such a classifier since the maximum-mass constraint is satisfied for all values of $c_1$ and $c_3$ by construction (because the density dependence of the symmetry energy for the 2-parameter model is stiff enough to support 2 $M_\odot$ stars).\\

\noindent \textbf{Bayesian analysis of GW170817 and X-ray pulse-profile data}

To obtain the posterior $P(\params|\data)$ presented in Fig.~\ref{fig:Present}, we use Bayes' theorem:
\begin{equation}
    P(\params|\data) \propto  \bigg[ \mathcal{L}(d_\textrm{J0030}|\params) \mathcal{L}(d_\textrm{J0740}|\params) \mathcal{L}(d_\textrm{J0437}|\params) \mathcal{L}(d_\textrm{GW170817}|\params) \bigg] \pi(\params)\,.
    \label{eq:bayes}
\end{equation}
Here, $\params$ is a set of EOS parameters, $\pi(\params)$ is the corresponding prior, and the likelihood function is obtained as the product of the likelihoods for different sources of data $\data$: the NICER observations of PSRs J0030, J0740, and J0437, as well as the gravitational-wave observation GW170817. 
We do not consider the event GW190425 given its low SNR~\autocite{LIGOScientific:2020aai} and negligible impact on the EOS (see, e.g., Dietrich et al.~\autocite{Dietrich:2020efo}).  

As priors we use uncorrelated uniform distributions for all EOS parameters.
For $c_1$ and $c_3$, the uniform priors span the range between $0$ and twice the laboratory value obtained using the Roy-Steiner analysis~\autocite{Siemens:2016jwj}.
For the speed-of-sound parameters, the uniform priors span the range between $0$ and the speed of light $c$. 
Before analyzing the gravitational-wave and NICER data, we additionally impose the constraint $M_\textrm{TOV} \geq 2 M_\odot$ as a hard cut, i.e., $\pi(\params) \propto \Theta(M_\textrm{TOV}(\params)-2M_\odot)$, where $\Theta$ denotes a step function. 
This prior constraint accounts for the radio observations of heavy pulsars with masses around $2M_\odot$~\autocite{Antoniadis:2013pzd,NANOGrav:2019jur,Fonseca:2021wxt} and is implemented implicitly when training the neural-network emulator for the stellar-structure equations.

In order to evaluate Eq.~\eqref{eq:bayes}, we first sample the distribution $\mathcal{L}(d_\textrm{GW170817}|\params) \pi(\params)$ using MCMC\autocite{emcee_pasp}.
The relevant likelihood is given by 
\begin{equation}
    \mathcal{L}(d_\textrm{GW170817}|\params) = \int dm_1 dm_2 \mathcal{L}(d_\textrm{GW170817}|m_1,m_2,\Lambda_1(m_1,\params),\Lambda_2(m_2,\params)) \pi(m_1,m_2)\,.
\end{equation}
The prior on the component masses is taken to be uniform in the 1--2 $M_\odot$ range.
We do not perform a parameter estimation for GW170817 starting from the GW strain data, as sampling over the LEC parameters would dramatically increase the computational cost of such an analysis, and instead take the likelihood $\mathcal{L}(d_\textrm{GW170817}|m_1,m_2,\tilde{\Lambda})$---where $\tilde{\Lambda}$ is the binary tidal deformability---to be proportional to the posterior computed in Abbott et al.\autocite{LIGOScientific:2018hze}. 
We use a Gaussian kernel density estimator (KDE) to construct the probability density function from the samples calculated in Abbott et al.\autocite{LIGOScientific:2018hze}. The use of a KDE might introduce uncertainties as the probability density function is not always reconstructed accurately by this method. However, we did not see any significant change in our results when the bandwidth of the KDE was varied to within 30\%. This is consistent with the general findings in the literature that KDE-induced uncertainties are sub-dominant for events that are only moderately informative, such as GW170817~\autocite{Kashyap:2025cpd,LandryEssick2020}.  
At each iteration of the MCMC sampler, the factor $\Theta(M_\textrm{TOV}(\params)-2M_\odot)$ is evaluated by the neural network classifier and the dependence of the binary tidal deformability $\tilde{\Lambda}$ on $\params$ in the above equation is determined by the emulator of the TOV equations.

\begin{figure}[t]
    \centering
    \includegraphics[width=0.65\textwidth]{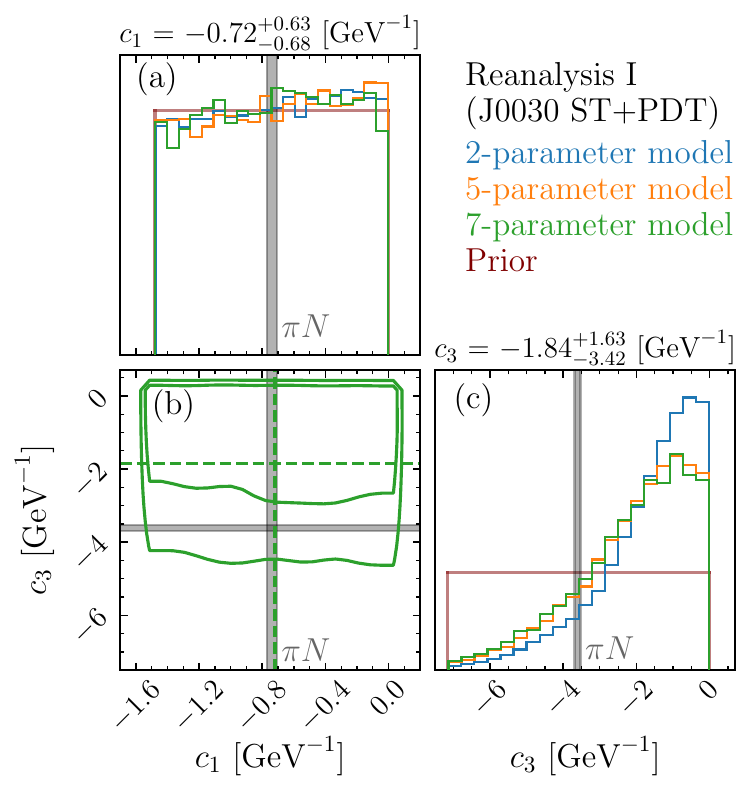}
    \caption{\textbf{Updated constraints from NICER Reanalysis I}. Panel (a) shows the posterior distribution function of the LEC $c_1$. Panel (c) shows the posterior distribution function of the LEC $c_3$. 
    Panel (b) depicts the correlation between $c_1$ and $c_3$ and displays iso-probability contours at the $68\%$ and $90\%$ confidence levels.
    These distributions are obtained by applying constraints from GW170817 and results from the recent reanalysis of NICER X-ray observations of PSR J0030~\autocite{Vinciguerra:2023qxq}.  
    Here, we use the mode with mass and radius of about $[1.4M_{\odot},11.5~\textrm{km}]$, i.e., using the hot-spot model ST+PDT. Source data for this figure are provided as a Source Data file.
    }
    \label{fig:Present_new}
\end{figure}

\begin{figure}[t]
    \centering
    \includegraphics[width=0.65\textwidth]{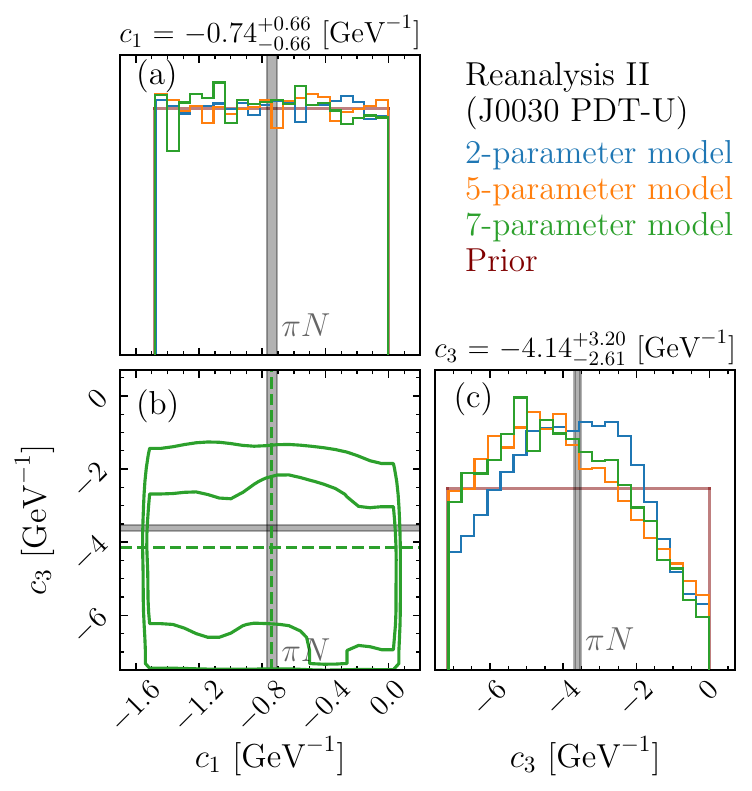}
    \caption{\textbf{Updated constraints from NICER Reanalysis II}.     Panel (a) shows the posterior distribution function of the LEC $c_1$. Panel (c) shows the posterior distribution function of the LEC $c_3$. 
    Panel (b) depicts the correlation between $c_1$ and $c_3$ and displays iso-probability contours at the $68\%$ and $90\%$ confidence levels. 
    These distributions are obtained by applying constraints from GW170817 and results from the recent reanalysis of NICER X-ray observations of PSR J0030~\autocite{Vinciguerra:2023qxq}.  
    Here, we use the mode with mass and radius of about $[1.7M_{\odot},14.5~\textrm{km}]$, i.e., using the hot-spot model PDT-U. Source data for this figure are provided as a Source Data file.
    }
    \label{fig:Present_new3}
\end{figure}

To account for the NICER observations, the samples obtained from the MCMC sampling above are given a weight proportional to $\prod_i \mathcal{L}(d_i|\params)$, where $i$ indexes the pulsar.
The individual NICER likelihoods are given as 
\begin{equation}
    \mathcal{L}(d_i|\params) = \int dM \mathcal{L}(d_i|M,R(M,\params)) \pi(M)\,,
\end{equation}
where the prior $\pi(M)$ is taken to be uniform between 1 $M_\odot$ and the maximal TOV mass $M_\textrm{TOV}$. The function $R(M,\params)$ is determined by solving the TOV equations using a high-fidelity solver.   
We approximate the likelihood functions $\mathcal{L}(d_i|M,R)$ to be proportional to the $M$-$R$ posteriors for PSRs~J0030\autocite{Riley:2019yda}, J0740\autocite{Salmi:2022cgy}, and J0437\autocite{Choudhury:2024xbk} respectively, and we use the hot-spot models corresponding to the main result in each publication. 
Note that this approximation is not strictly valid for J0740 because a Gaussian prior on the pulsar's mass, informed by radio observations, was used to obtain the $M$-$R$ posterior in Salmi et al.\autocite{Salmi:2022cgy} 
Consequently, the existence of heavy pulsars is effectively accounted for twice in our analysis: first by applying a hard cut in $\pi(\params)$ and then by the term $\mathcal{L}(d_{J0740}|M,R)$. 
However, since the Gaussian prior on the pulsar's mass used in Salmi et al.\autocite{Salmi:2022cgy} lies above 2 $M_\odot$ at the $68\%$ CL and our hard-cut term is constant in this region, we conclude that the double counting of the pulsar's mass measurement has only a minor effect on our results.
Our main results for the marginal posteriors on $c_1$ and $c_3$ obtained in this manner are shown in Fig.~\ref{fig:Present}. 

Recently, the NICER collaboration published a reanalysis of PSR~J0030 using additional data~\autocite{Vinciguerra:2023qxq}. 
In Figs.~\ref{fig:Present_new} and~\ref{fig:Present_new3}, we present results on $c_1$ and $c_3$ using two results of this reanalysis: one with the hot-spot model ST+PDT with mass and radius of about $[1.4M_{\odot},11.5~\textrm{km}]$ (reanalysis I) and the other with the hot-spot model PDT-U with a mean at about $[1.7M_{\odot},14.5~\textrm{km}]$ (reanalysis II).
While the results of reanalysis I are in agreement with Fig.~\ref{fig:Present}, reanalysis II leads to a notable difference in the posterior on $c_3$. 
However, when considering the 90\% CL given in Fig.~\ref{fig:convergence} for the present astrophysical data, this only changes the allowed $c_3$ range from $[-6.15,-0.31]$~GeV$^{-1}$ to $[-6.75,-0.94]$~GeV$^{-1}$. \\

\noindent \textbf{Hierarchical analysis of next-generation events}

For our study of the impact of next-generation gravitational-wave networks, we analyze the $N$ loudest simulated events in a hierarchical approach using Bayes' theorem~\autocite{LandryEssick2020,Landry:2022rxu}, i.e.,
\begin{equation}
    P(\params|d) \propto \bigg[ \prod_{i=1}^{N} \int dm_{1,i} dm_{2,i} \mathcal{L}(d_i|m_{1,i},m_{2,i},\tilde{\Lambda}_i(m_{1,i},m_{2,i},\params)) \pi(m_{1,i},m_{2,i}) \bigg] \pi(\params)\,.
    \label{eq:bayes_NG}
\end{equation}
In such an approach, EOS parameters are simultaneously informed by all of the events and the full multidimensional posterior is calculated to properly account for correlations between the various inferred parameters.
While one in principle also has to infer the parameters of the source population simultaneously with $\params$, we here circumvent this issue by choosing the priors on the component masses $\pi(m_1,m_2)$ to be the same as the distribution used to generate the population of events, i.e., a uniform distribution in the range 1--2 $M_\odot$ for both $m_1$ and $m_2$. 
Because the population model is fixed and independent of the EOS, the GW selection bias towards detecting more massive systems has no impact on the recovery of the EOS parameters and can be neglected~\autocite{LandryEssick2020}.

The single-event likelihoods $\mathcal{L}(d_i|m_{1,i},m_{2,i},\tilde{\Lambda}_i)$ are computed within the Fisher matrix approximation~\autocite{Cutler:1994ys,Poisson:1995ef,Borhanian:2020ypi}, which is valid for high-SNR events, using the TaylorF2 waveform model~\autocite{BuonannoIyer2009,WadeCreighton2014} and the next-generation detector network described in the main text. 
This results in uncorrelated Gaussian likelihoods on the chirp masses, symmetric mass ratios, and the binary tidal deformabilities. 
Given that the quietest event in our sample has an SNR of about 225, we do not expect the results of our inference---performed at the level of a population of events---to be impacted significantly by the Fisher matrix approximation. 
In order to verify this, we re-computed all 20 single-event likelihoods (proportional to the single-event posteriors) with Bayesian parameter estimation~\autocite{Veitch:2014wba}. 
For each event, a zero-noise injection was created and the binary parameters were recovered using the heterodyne likelihood model~\autocite{Finstad:2020sok} implemented in~\textsc{PyCBC}~\autocite{Biwer:2018osg}. 
In Fig.~\ref{fig:convergence}, we see that the results obtained using the Fisher matrix approximation (circle) are very similar to those obtained using parameter estimation on zero-noise injections (diamond) for the 7-parameter model at $N=20$.
This indicates the validity of employing the Fisher matrix approximation for the $20$ single-event likelihoods.

As in the analysis of GW170817 and the NICER data, the prior $\pi(\params)$ is taken to be uniform on all the EOS parameters, along with the constraint that $\pi(\params) \propto \Theta(M_\textrm{TOV}(\params)-2M_\odot)$.
The posterior distribution defined in Eq.~\eqref{eq:bayes_NG} is sampled using MCMC for different values of $N$, with the relation between $\tilde{\Lambda}$ and $\params$ in all $20$ single-event likelihoods being evaluated by the neural network emulator. 

In Fig.~\ref{fig:convergence}, our results for the three EOS models are shown for $N=1, 3, 5, 10, 15,$ and $20$. 
The marginal posteriors on $c_3$ are obtained by integrating Eq.~\eqref{eq:bayes_NG} over all EOS parameters except $c_3$. 
We average over a large number of orderings of our $N$ events, i.e.,  for a given $N$, the MCMC sampling of Eq.~\eqref{eq:bayes_NG} is repeated for $100$ different permutations of $N$ events drawn from our pool of $20$ events.
Upon performing this average, we find that the statistical uncertainties of the posteriors approximately obey a $1/\sqrt{N}$ behavior, similar to the findings of Landry et al.\autocite{LandryEssick2020} and Kunert et al.\autocite{Kunert:2021hgm}.

We found that, even at $N=20$, the marginal posterior on $c_1$ is uninformative and is fully determined by the prior in the case of the 5- and 7-parameter models. 
This highlights that $c_1$ cannot be constrained from astrophysical data even with next-generation detectors given the lack of sensitivity of the EOS of neutron matter to $c_1$. 
For the 2-parameter model, we found a small but noticeable shift of the posterior towards smaller values of $c_1$. 
However, this is a consequence of the limitations of the 2-parameter model and further highlights the importance of the marginalization over the high-density EOS. 

\begin{figure}[t]
    \centering
    \includegraphics[width=0.65\textwidth]{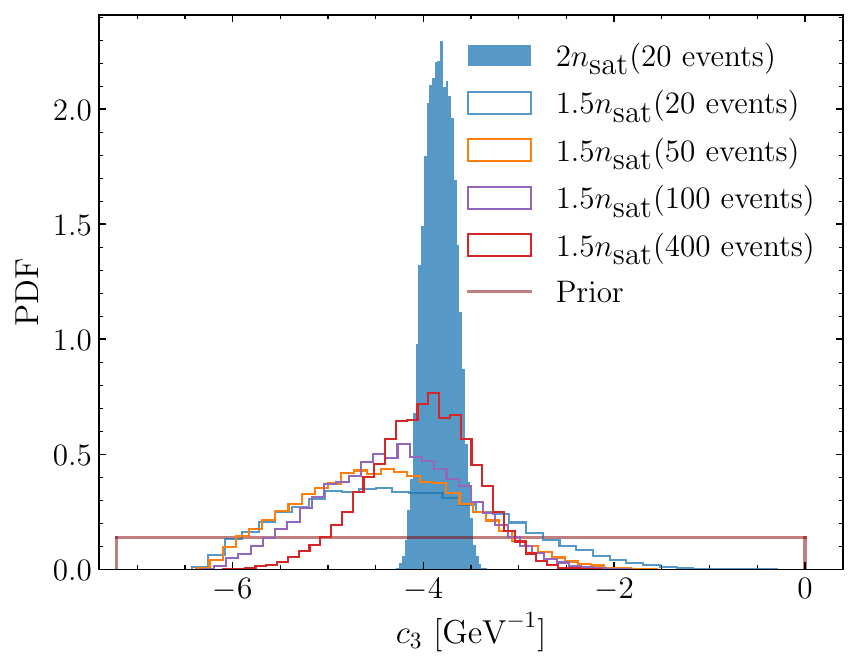}
    \caption{\textbf{Results with lower EFT breakdown scale}. A re-analysis of our projected constraints for $c_3$ in the era of third-generation GW detectors, but assuming validity of the EFT expansion up to $1.5n_\textrm{sat}$ rather than $2 n_\textrm{sat}$ with different number of events. Source data for this figure are provided as a Source Data file.}
    \label{fig:15nsat}
\end{figure}

Finally, we note that our constraints on $c_3$ are sensitive to the assumed value of the EFT breakdown density. 
In Fig.~\ref{fig:15nsat}, we have re-computed our constraints by decreasing the breakdown density from $2n_\textrm{sat}$ to $1.5n_\textrm{sat}$. 
The employed high-density model is similar to the 7-parameter model, but with an additional parameter for the squared sound speed at $2n_\textrm{sat}$. 
We find that, when only $20$ events are used, the marginal posterior on $c_3$ significantly broadens, compared to the case with a breakdown density of $2n_\textrm{sat}$. 
However, the constraint tightens as more events are added. 
When analyzing almost all 400 events observed within one year, we find that the posterior shows a remarkable convergence towards the $2n_\textrm{sat}$ result. 
This indicates that a lower value of the EFT breakdown density still allows for significant constraints on $c_3$, only with more events required, and that no major qualitative shift in EFT behavior occurs between the two breakdown densities.
This approach can potentially allow us to constrain the breakdown density in future work. \\

\noindent \textbf{Data availability}

The posterior samples for GW170817 used in our analysis can be found at \url{https://dcc.ligo.org/LIGO-P1800061/public}\autocite{abbott_gw170817_data}. 
The mass-radius posteriors from the NICER analyses are available at~\url{https://zenodo.org/records/3386449}\autocite{riley_2019_3386449} (PSR J0030, 2019), \url{https://zenodo.org/records/8239000}\autocite{vinciguerra_2023_8239000} (PSR J0030, 2024), 
\url{https://zenodo.org/records/6827537}\autocite{salmi_2022_6827537} (PSR J0740),
and \url{https://zenodo.org/records/12703175}\autocite{choudhury_2024_12703175} (PSR J0437). 
The datasets generated in this study, including all posterior samples and emulator training/validation data, have been deposited in this~\href{https://doi.org/10.5281/zenodo.16878207}{Zenodo repository}~\autocite{somasundaram_zenodo}. Source data are provided with this paper. \\

\noindent \textbf{Code availability}

The package \textsc{gwbench}~\autocite{Borhanian:2020ypi} used to construct the Fisher matrix likelihoods can be found at \url{https://gitlab.com/sborhanian/gwbench}. 
The injection and parameter estimation-based recovery runs were performed using \textsc{PyCBC}~\autocite{Biwer:2018osg}, a publicly available software package that can be found at \url{https://github.com/gwastro/pycbc}.  
All codes developed for the analyses in this paper can be found in this~\href{https://github.com/svisak/multimessenger_3N_constraints}{GitHub repository}~\autocite{somasundaram_zenodo_cde}.

\printbibliography

\subsection*{Acknowledgements}

We thank C.L.~Armstrong, K.~Godbey, and P.~Giuliani for feedback about the implementation of the PMM. 
We further thank D.~Brown, C.~Capano, C. Forssén, K.~Hebeler, and W.G. Jiang for useful discussions. 
We also thank the Institute for Nuclear Theory at the University of Washington for its kind hospitality and stimulating research environment.

R.S. acknowledges support from the Nuclear Physics from Multi-Messenger Mergers (NP3M) Focused Research Hub which is funded by the National Science Foundation under Grant Number 21-16686, and from the Laboratory Directed Research and Development program of Los Alamos National Laboratory under project number 20220541ECR.
I.S. and A.S. were supported in part by the European Union's Horizon 2020 research and innovation programme (Grant Agreement No.~101020842).
S.D., A.E.D., and I.T. were supported by the Laboratory Directed Research and Development program of Los Alamos National Laboratory under project number 20230315ER.
A.E.D. was also supported in part by the U.S. Department of Energy, Office of Science, Office of Workforce Development for Teachers and Scientists (WDTS) under the Science Undergraduate Laboratory Internships Program (SULI).
S.D. was also supported by the Laboratory Directed Research and Development program of Los Alamos National Laboratory project 20250750ECR.
I.T. was also supported by the U.S. Department of Energy, Office of Science, Office of Nuclear Physics, under contract No.~DE-AC52-06NA25396, by the U.S. Department of Energy, Office of Science, Office of Advanced Scientific Computing Research, Scientific Discovery through Advanced Computing (SciDAC) NUCLEI program, and by the Laboratory Directed Research and Development program of Los Alamos National Laboratory under project numbers 20220541ECR.
Y.D. was supported by the Deutsche Forschungsgemeinschaft (DFG, German Research Foundation) – Project-ID 279384907 – SFB 1245.
P.L. is supported by the Natural Sciences \& Engineering Research Council of Canada (NSERC). Research at Perimeter Institute is supported in part by the Government of Canada through the Department of Innovation, Science and Economic Development and by the Province of Ontario through the Ministry of Colleges and Universities.
Computational resources have been provided by the Los Alamos National Laboratory Institutional Computing Program, which is supported by the U.S. Department of Energy National Nuclear Security Administration under Contract No. 89233218CNA000001, and by the National Energy Research Scientific Computing Center (NERSC), which is supported by the U.S. Department of Energy, Office of Science, under contract No. DE-AC02-05CH11231, using NERSC awards NP-ERCAP-m3319 and NP-ERCAP-m4338.
I.S. and A.S. gratefully acknowledge the computing time provided on the high-performance computer Lichtenberg II at the TU Darmstadt. This is funded by the German Federal Ministry of Education and Research (BMBF) and the State of Hesse. I.S. and A.S. also gratefully acknowledge the Gauss Centre for Supercomputing e.V. (www.gauss-centre.eu) for funding this project by providing computing time on the GCS Supercomputer JUWELS~\cite{JUWELS} at Jülich Supercomputing Centre (JSC). This document has been approved for unlimited release, and was assigned LA-UR-24-29187. 

\subsection*{Author contributions}
R. Somasundaram and I. Svensson share first authorship. \\
Conceptualization: RS, IS, PL, AS, IT; \\
Methodology: RS, IS, SD, PL, AS, IT; \\
Data curation: RS, IS, SD, PL; \\
Software: RS, IS, SD, AED, YD, PL; \\
Validation: RS, IS, SD, AED, YD, PL, AS, IT; \\
Formal analysis: RS, IS, SD, AED, PL, AS, IT; \\
Resources: IS, SD, PL, AS, IT; \\
Funding acquisition: SD, PL, AS, IT; \\
Project administration: RS, IS, PL, AS, IT; \\
Supervision: RS, AS, IT; \\
Visualization: RS, IS, AED, PL; \\
Writing–original draft: RS, IS, YD, PL, AS, IT; \\
Writing–review and editing: RS, IS, SD, AED, YD, PL, AS, IT.

\subsection*{Competing interests} 

The authors declare no competing interests.

\end{document}